\input epsf

\documentstyle{multi}
\begin{document}
\pagenumbering{arabic}
\setcounter{page}{1}

\chapterb{The Gauss linking number in quantum gravity}
{Rodolfo Gambini}{Intituto de F\'{\i}sica, Facultad de
Ciencias, \\Tristan Narvaja 1674, Montevideo, Uruguay} {Jorge
Pullin}{Center for Gravitational Physics and Geometry,\\
The Pennsylvania State University, University Park, PA 16802, USA}

\authormark{Rodolfo Gambini and Jorge Pullin}

{\tt To appear in ``Knots and quantum gravity'', J. Baez editor, 
Oxford University Press (1993). }
\section*{Abstract}
\vspace{-8cm} 
\begin{flushright}
\baselineskip=15pt
CGPG-93/10-1  \\
gr-qc/9310025 \\
October 18th 1993\\
\end{flushright}
\vspace{6.5cm}

We show that the exponential of the Gauss (self) linking number of a
knot is a solution of the Wheeler-DeWitt equation in loop space with a
cosmological constant. Using this fact, it is straightforward to prove
that the second coefficient of the Jones Polynomial is a solution of the
Wheeler-DeWitt equation in loop space with no cosmological constant. We
perform calculations from scratch, starting from the connection
representation and give details of the proof. Implications for the
possibility of generation of other solutions are also discussed.

\section{Introduction}

The introduction of the loop representation for quantum gravity has made
it possible for the first time to find solutions to the Wheeler-DeWitt
equation (the quantum Hamiltonian constraint) and therefore to have
possible candidates to become physical wavefunctions of the
gravitational field. In the loop representation the Hamiltonian
constraint has nonvanishing action only on functions of intersecting
loops. It was first argued that by considering wavefunctions with
support on smooth loops one could solve the constraint straightforwardly
\cite{JaSm,RoSm}. However it was later realized that such solutions are
associated with degenerate metrics (metrics with zero determinant) and
this posed inconsistencies if one wanted to couple the theory
\cite{BrPu}. For instance if one considered general relativity with a
cosmological constant it turns out that nonintersecting loop states also
solve the Wheeler-DeWitt equation for arbitrary values of the
cosmological constant. This does not appear as reasonable since
different values of the cosmological constant lead to widely different
behaviors in general relativity.

Therefore the problem of solving the Wheeler-DeWitt equation in loop
space is far from solved and has to be tackled by considering the action
of the Hamiltonian constraint in the loop representation on states based
on (at least triply-)intersecting loops, as depicted in figure 1.

\begin{figure}[t]
\epsffile{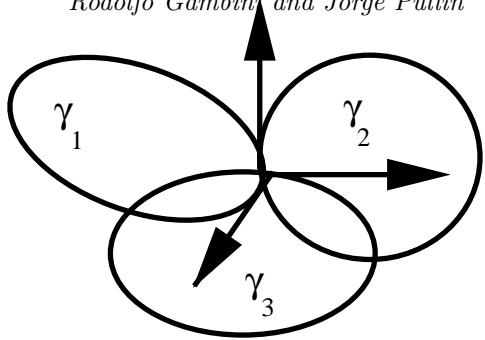}
\vspace{.3cm}
\caption{At least a triple intersection is needed to have a  
state associated with a nondegenerate 
three dimensional metric in the loop representation. And even in this
case the metric is nondegenerate only at the point of intersection}
\end{figure}

Although performing these kind of direct calculations is now possible,
since well defined expressions for the Hamiltonian constraint exist in
terms of the loop derivative \cite{Ga} (see also \cite{BrPu}) and
actually some solutions have been found with this
approach \cite{BrGaPu}, another line of reasoning has also proved to be
useful.

This other approach is based on the fact that the exponential of the
Chern-Simons term based on Ashtekar's connection,
\begin{equation}
\Psi^{CS}_\Lambda[A] = \exp \left( {-{\textstyle {6\over \Lambda}}}
\int d^3x (A_a^i \partial_b A_c^i +{\textstyle {2\over 3}}
A_a^i A_b^j A_c^k \epsilon^{ijk}) \epsilon^{abc} \right)
\end{equation}
is a solution of the Wheeler-DeWitt equation in the connection
representation with a cosmological term,
\begin{equation}
\hat{H} = \epsilon^{ijk} {\delta \over \delta A_a^i} {\delta \over
\delta A_b^j} F_{ab}^k + {\textstyle {\Lambda \over 6}} 
\epsilon^{ijk} \epsilon^{abc} {\delta \over \delta A_a^i} {\delta \over
\delta A_b^j} {\delta \over
\delta A_c^k}\label{Ham}
\end{equation}
(the first term is just the usual Hamiltonian constraint with
$\Lambda=0$ and the second term is just $\Lambda \; {\rm det g}$ where
${\rm det g}$ is the determinant of the spatial part of the metric
written in terms of triads.

To prove this fact one simply needs to notice that for the Chern-Simons
state introduced above,
\begin{equation}
{\delta \over
\delta A_a^i} \Psi^{CS}[A] = {\textstyle -{6 \over \Lambda}} \epsilon^{abc}
F_{bc}^i \Psi^{CS}[A]
\end{equation}
and therefore the rightmost functional derivative in the cosmological
constant term of the Hamiltonian (\ref{Ham}) produces a term in $F_{ab}^i$
that exactly cancels the contribution from the vacuum Hamiltonian
constraint. 

This result for the Chern-Simons state in the connection representation
has an immediate counterpart in the loop representation, since the
transform of the Chern-Simons state into the loop representation,
\begin{equation}
\Psi^{CS}_\Lambda(\gamma)= \int dA\; \Psi^{CS}_\Lambda[A] \;W_\gamma[A]
\end{equation}
can be interpreted as the expectation value of the Wilson loop in a
Chern-Simons theory,
\begin{equation}
\Psi^{CS}_\Lambda(\gamma)= \int dA\; e^{-{\textstyle{6\over\Lambda}} S_{CS}}
\;
W_\gamma[A] = <W_\gamma> 
\end{equation}
with coupling constant ${6\over \Lambda}$, 
and we know due to the insight of Witten \cite{Wi} that this coincides
with the Kauffman bracket of the loop. Therefore the state
\begin{equation}
\Psi^{CS}(\gamma) = {\rm Kauffman\; Bracket}_{\Lambda}(\gamma)
\end{equation}
should be a solution of the Wheeler-DeWitt equation in loop space. 

This last fact can actually be checked in a direct fashion using the
expressions for the Hamiltonian constraint in loop space of reference
\cite{Ga}. In order to do this we make use of the identity,
\begin{equation}
{\rm Kauffman\; Bracket}_{\Lambda}(\gamma) = e^{\Lambda {\rm
Gauss}(\gamma)} {\rm Jones\; Polynomial}_\Lambda(\gamma)
\end{equation}
which relates the Kauffman Bracket, the Gauss (self) linking number
and the Jones Polynomial. The Gauss self linking number is framing
dependent, and so is the Kauffman Bracket. The Jones Polynomial
however, is framing independent. This raises the issue of up to what
extent statemets about the Kauffman Bracket being a state of gravity
are valid since one expects states of quantum gravity to be truly
diffeomorphism invariant objects and framing is always dependent on an
external device which should conflict with the diffeomorphism
invariance of the theory. Unfortunately it is not clear at present how
to settle this issue since it is tied to the regularization procedures
used to define the constraints. We will return to these issues in the
final discussion.

We now expand both the Jones Polynomial and the exponential of the Gauss
linking number in terms of $\Lambda$ and get the expression,
\begin{eqnarray}
{\rm Kauffman\;Bracket}_\Lambda(\gamma) &=& 1 + 
{\rm Gauss}(\gamma) \Lambda +\\
&+&
({\rm Gauss}(\gamma)^2 +a_2(\gamma)) \Lambda^2 +\nonumber\\
&+&({\rm Gauss}(\gamma)^3 +
{\rm Gauss}(\gamma)^2 a_2(\gamma) +a_3(\gamma)) \Lambda^3 +\nonumber\\
&+&\ldots\nonumber
\label{expansion}
\end{eqnarray}
where $a_2, a_3$ are the second and third coefficient of the infinite
expansion of the Jones Polynomial in terms of $\Lambda$. $a_2$ is known
to coincide with the second coefficient of the Conway polynomial
\cite{GuMaMi}. 

One could now apply the Hamiltonian constraint in the loop
representation with a cosmological constant to the expansion
(\ref{expansion}) and one would find that certain conditions have to be
satisfied if the Kauffman Bracket is to be a solution. Among them it was
noticed \cite{BrGaPuessay} that
\begin{equation}
\hat{H}_0 a_2(\gamma)=0
\end{equation}
where $\hat{H}_0$ is the Hamiltonian constraint without cosmological
constant (more recently it has also been shown that $\hat{H}_0
a_3(\gamma)=0$ \cite{Gr} but we will not discuss it here). 

Summarizing, the fact that the Kauffman Bracket is a solution of the
Wheeler-DeWitt equation with cosmological constant seems to have as a
direct consequence that the coefficients of the Jones Polynomial are
solutions of the vacuum ($\Lambda=0$) Wheeler-DeWitt equation!. This
partially answers the problem of framing we pointed out above. Even if
one is reluctant to accept the Kauffman Bracket as a state because of
its framing dependence, it can be viewed as an intermediate step of a
framing-dependent proof that the Jones Polynomial (which is a framing
independent invariant) solves the Wheeler-DeWitt equation (it should be
stressed that we only have evidence that the first two nontrivial
coefficients are solutions).

The purpose of this paper is to present a rederivation of these facts
from a different, and to our understanding simpler, perspective. We will
show that the (framing-dependent) knot invariant,
\begin{equation}
\Psi^G_\Lambda(\gamma) = e^{\Lambda {\rm Gauss}(\gamma)}
\end{equation}
is also a solution of the Wheeler-DeWitt equation with cosmological
constant. It can be viewed as an "Abelian limit" of the Kauffman Bracket
(more on this in the conclusions). Given this fact, one can therefore
consider their difference divided by $\Lambda^2$,
\begin{equation}
D_\Lambda(\gamma)= {({\rm Kauffman\; Bracket}_\Lambda(\gamma)
-\Psi^G_\Lambda(\gamma))\over \Lambda^2}
\end{equation}
which is also a solution of the Wheeler-DeWitt equation with a
cosmological constant. This difference is of the form,
\begin{equation}
D_\Lambda(\gamma) = a_2(\gamma)+ (a_3(\gamma) + {\rm
Gauss}(\gamma)^2 a_2) \Lambda +\ldots
\end{equation}

Now, this difference is a state for all values of $\Lambda$, in
particular, for $\Lambda=0$. This means that $a_2(\gamma)$ should be a
solution of the Wheeler-DeWitt equation. This confirms the proof given
in references \cite{BrGaPu,BrGaPuessay}. 

Therefore we see that by noticing that the Gauss linking number is a
state with cosmological constant, it is easy to prove that the second
coefficient of the infinite expansion of the Jones Polynomial (which
coincides with the second coefficient of the Conway Polynomial) is a
solution of the Wheeler-DeWitt equation with $\Lambda=0$.

The rest of this paper will be devoted to a detailed proof that the
exponential of the Gauss linking number solves the Wheeler-DeWitt
equation with cosmological constant.  To this aim we will derive
expressions for the Hamiltonian constraint with cosmological constant in
the loop representation.  We will perform the calculation explicitly for
the case of a triply-self intersecting loop, the more interesting case
for gravity purposes (it should be noticed that all the arguments
presented above were independent of the number and order of
intersections of the loops, we just present the explicit proof for a
triple intersection since in three spatial dimensions it represents the
most generic type of intersection). 

Apart from presenting this new state, we think the calculations
exhibited in this paper should help the reader get into the details of
how these calculations are performed and make an intuitive contact
between the expressions in the connection and the loop representation. 

In section 2 we derive the expression of the Hamiltonian constraint
(with a cosmological constant) in the loop representation for a triply
intersecting loop in terms of the loop derivative. In section 3 we
write an explicit analytic expression for the Gauss linking number and
prove that it is a state of the theory. We end in section 4 with a
discussion of the results.

\section{The Wheeler-DeWitt equation in terms of loops}

Here we derive the explicit form in the loop representation of the
Hamiltonian constraint with a cosmological constant. The derivation
proceeds along the following lines. Suppose one wants to define the
action of an
operator $\hat{O}_L$ on a wavefunction in the loop representation
$\Psi(\gamma)$. Applying the transform,
\begin{equation}
\hat{O}_L \Psi(\gamma) \equiv \int dA \hat{O}_L W_\gamma[A] \Psi[A]
\end{equation}
the operator $\hat{O}$ in the right member acts on the loop dependence
of the Wilson loop. On the other hand, this definition should agree
with,
\begin{equation}
\hat{O}_L = \int dA W_\gamma[A] \hat{O}_C \Psi[A]
\end{equation}
where $\hat{O}_C$ is the connection representation version of the
operator in question. Therefore, it is clear 
that,
\begin{equation}
\hat{O}_L W_\gamma[A] \equiv \hat{O}^\dagger_C W_\gamma[A]
\end{equation}
where $\dagger$ means the adjoint operator with respect to the measure
of integration $dA$. If one assumes that the measure is trivial, the
only effect of taking the adjoint is to reverse the factor ordering of
the operators. 

Concretely, in the case of the Hamiltonian constraint (without
cosmological constant)
\begin{equation}
\hat{H}_C = \epsilon^{ijk} {\delta \over \delta A_a^i} {\delta \over
\delta A_b^j} F_{ab}^k 
\end{equation}
and therefore
\begin{equation}
\hat{H}_L W_\gamma[A] \equiv 
\epsilon^{ijk} F_{ab}^k  {\delta \over \delta A_a^i} {\delta
\over \delta A_b^j} W_\gamma[A].
\end{equation}

We now need to compute this quantity explicitly. For that we need the
expression of the functional derivative of the Wilson loop with respect
to the connection,
\begin{equation}
{\delta \over \delta A_a^i(x)} W_\gamma[A] = \oint dy^a \delta(y-x) Tr[
{\rm Pexp}(\int_o^y dz^b A_b ) \tau^i {\rm Pexp}(\int_y^o dz^b A_b) ]
\end{equation}
where $o$ is the basepoint of the loop. The Wilson loop is therefor
"broken" at the point of action of the functional derivative and a Pauli
matrix ($\tau^i$) is inserted. It is evident that with this action of
the functional derivative the Hamiltonian operator is not well defined.
We need to regularize it,
\begin{equation}
\hat{H}_L W_\gamma[A] = \lim_{\epsilon\rightarrow0} f_\epsilon(x-z)
\epsilon^{ijk} F_{ab}^k(x)  {\delta \over \delta A_a^i(x)} {\delta
\over \delta A_b^j(z)} W_\gamma[A].
\end{equation}
where $f_{\epsilon}(x-z)\rightarrow \delta(x-z)$ when
$\epsilon\rightarrow 0$.  Caution should be exercised, since such
point-splitting breaks the gauge invariance of the Hamiltonian.  There
are a number of ways of fixing this situation in the language of
loops.  One of them is to define the Hamiltonian inserting pieces of
holonomies connecting the points $x$ and $z$ between the functional
derivatives to produce a gauge invariant quantity \cite{JaSm}.  Here
we will only study the operator in the limit in which the regulator is
removed, therefore we will not be concerned with these issues.  A
proper calculation would require their careful study.

It is immediate from the above definitions that the Hamiltonian
constraint vanishes in any regular point of the loop, since it yields
a term $dy^a dy^b F_{ab}^i$ which vanishes due to the antisymmetry of
$F_{ab}^i$ and the symetry of $dy^a dy^b$ at points where the loop is
smooth. However, at intersections there can be nontrivial
contributions. Here we compute the contribution at a point of triple
self-intersection,
\begin{eqnarray}
&&
\epsilon^{ijk} F_{ab}^k(x)  {\delta \over \delta A_a^i(x)} {\delta
\over \delta A_b^j(z)} W_{\gamma_1\circ\gamma_2\circ\gamma_3}[A] =
\epsilon^{ijk} F_{ab}^k(x)\times\\
&&
\times(
\dot{\gamma}^a_1 \dot{\gamma}^b_2 W_{\gamma_1 \tau^j \gamma_2 \tau^k
\gamma_3}[A]+
\dot{\gamma}^a_1 \dot{\gamma}^b_3 
W_{\gamma_1 \tau^j \gamma_2 \gamma_3 \tau^k
}[A]+
\dot{\gamma}^a_2 \dot{\gamma}^b_3 
W_{\gamma_1 \gamma_2 \tau^j \gamma_3 \tau^k
}[A])\nonumber
\end{eqnarray}
where we have denoted $\gamma=\gamma_1\circ\gamma_2\circ\gamma_3$
where $\gamma_i$ are the "petals" of the loop as indicated in figure
1.  By $W_{\gamma_1 \tau^j \gamma_2 \tau^k \gamma_3}[A]$ we really
mean take the holonomy from the basepoint along $\gamma_1$ up to just
before the intersection point, insert a Pauli matrix, continue along
$\gamma_1$ to the intersection, continue along $\gamma_2$ and just
before the intersection insert another Pauli matrix, continue to the
intersection and complete the loop along $\gamma_3$. One could pick
"after" the intersection instead of "before" to include the Pauli
matrices and it would make no difference since we are concentrating in
the limit in which the regulator is removed in which the insertions
are done at the intersection. By $\dot{\gamma}_1^a$ we mean the
tangent to the petal number 1 just before the intersection (where the
Pauli matrix was inserted). This is just a shorthand for expressions
like $\oint dy^a
\delta(x-y)$ when the point $x$ is close to the intersection, so
strictly speaking $\dot{\gamma}^a_1$ really is a distribution that is
nonvanishing only at the point of intersection.

One now uses the following identity for traces of $SU(2)$ matrices,
\begin{equation}
\epsilon^{ijk} W_{\alpha \tau^j \beta \tau^k }[A] = {\textstyle
{1\over 2}} (W_{\alpha \tau^i}[A] 
W_{\beta}[A] - W_{\alpha}[A] W_{\beta \tau^i}[A])
\end{equation}
which is a natural generalization to the case of loops with insertions 
of the $SU(2)$ Mandelstam identities,
\begin{equation}
W_{\alpha}[A] W_{\beta}[A] = W_{\alpha \beta}[A] + W_{\alpha
\bar{\beta}}[A]
\end{equation}
where $\bar{\beta}$ means the loop opposite to  $\beta$. The result of
the application of these identities to the expression of the Hamiltonian
is, 
\begin{eqnarray}
&&\hat{H} W_\gamma[A] = \\
&&{\textstyle {1 \over 2}} F_{ab}^i
\left(
\dot{\gamma}_1^a \dot{\gamma}_2^b W_{\bar{\gamma_2} \gamma_3
\gamma_1 \tau^i}[A] 
-\dot{\gamma}_1^a \dot{\gamma}_3^b W_{\bar{\gamma_1} \gamma_2
\gamma_3 \tau^i}[A] +
\dot{\gamma}_1^a \dot{\gamma}_2^b W_{\bar{\gamma_3} \gamma_1
\gamma_2 \tau^i}[A]\right).\nonumber
\end{eqnarray}

This expression can be further rearranged making use of the loop
derivative. The loop derivative $\Delta_{ab}(\pi_o^x)$ 
\cite{GaTr,Ga} is the differentiation
operator that appears in loop space when one considers two loops to be
"close" if they differ by an infinitesimal loop appended through a path
$\pi_o^x$ going from the basepoint to a point of the manifold $x$ as
shown in figure 2. Its definition is
\begin{equation}
\Psi(\pi_o^x \delta \gamma \pi_x^o \gamma) = (1 +\sigma^{ab}
\Delta_{ab}(\pi_o^x)) \Psi(\gamma)
\end{equation}
where $\sigma^{ab}$ is the element area of the infinitesimal loop
$\delta \gamma$ and by $\pi_o^x \delta \gamma \pi_x^o \gamma$ we mean
the loop obtained by traversing the path $\pi$ from the basepoint to
$x$, the infinitesimal loop $\delta \gamma$, the path $\pi$ from $x$
to the basepoint and then the loop $\gamma$.

\begin{figure}
\epsffile{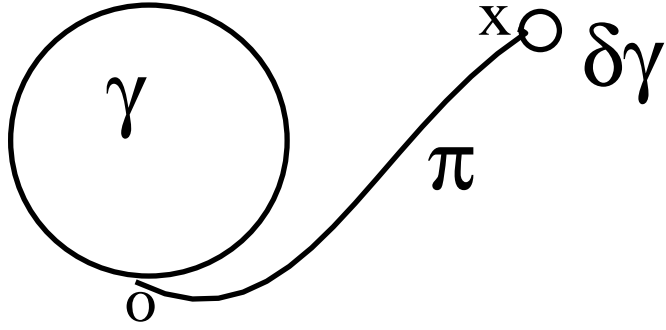}
\caption{The loop defining the loop derivative}
\end{figure}

We will not
discuss all its properties here. The only one we need is that the loop
derivative of a Wilson loop taken with a path along the loop is given
by, 
\begin{equation}
\Delta_{ab}(\gamma_o^x) W_\gamma[A] = {\rm Tr}[F_{ab}(x) {\rm Pexp}(\oint
dy^c A_c)]
\end{equation}
which reflects the intuitive notion that a holonomy of an infinitesimal
loop is related with the field tensor. Therefore we can write
expressions like,
\begin{equation}
F_{ab}^i W_{\bar{\gamma_2} \gamma_3
\gamma_1 \tau^i}[A] 
\end{equation}
as,
\begin{equation}
\Delta_{ab}(\gamma_1) W_{\bar{\gamma_2} \gamma_3
\gamma_1}[A] 
\end{equation}
and the final expression for the Hamiltonian constraint in the loop
representation can therefore be read off as follows,
\begin{eqnarray}
\hat{H} \Psi(\gamma) =&& {\textstyle {1 \over 2} }
(
\dot{\gamma}_1^a \dot{\gamma}_2^b 
\Delta_{ab}(\gamma_1) \Psi(\bar{\gamma_2} \gamma_3 \gamma_1) +\\
&&+\dot{\gamma}_1^a 
\dot{\gamma}_3^b \Delta_{ab}(\gamma_3) \Psi(\bar{\gamma_1}
\gamma_2 \gamma_3) +
\dot{\gamma}_1^a \dot{\gamma}_2^b 
\Delta_{ab}(\gamma_3) \Psi(\bar{\gamma_3} \gamma_1
\gamma_2 )).\nonumber
\end{eqnarray}

This expression could be obtained by particularizing that of refererence
\cite{Ga} to the case of a triple self-intersecting loop and rearranging
terms a bit using the Mandelstam identities. However we thought that a
direct derivation for this particular case would be useful for
pedagogical purposes.

We now have to find the loop representation form of the operator
corresponding to the determinant of the metric in order to represent the
second term in (\ref{Ham}). We proceed in a similar fashion, first
computing the action of the operator in the connection representation,
\begin{equation}
\hat{\rm det q} = \epsilon^{ijk} \epsilon_{abc} {\delta \over \delta
A_a^i} {\delta \over \delta A_b^j}{\delta \over \delta A_c^k}
\end{equation}
on a Wilson loop,
\begin{equation}
\epsilon^{ijk} \epsilon_{abc} {\delta \over \delta
A_a^i} {\delta \over \delta A_b^j}{\delta \over \delta A_c^k} W_{\gamma}[A]
= \epsilon_{abc} \epsilon^{ijk}
\dot{\gamma}_1^a\dot{\gamma}_2^b\dot{\gamma}_2^c \left(
W_{\gamma_1 \tau^i \gamma_2 \tau^j \gamma_3 \tau^k}[A]\right).
\end{equation}
This latter expression can be rearranged with the following identity
between holonomies with insertions of Pauli matrices,
\begin{equation}
\epsilon^{ijk} W_{\gamma_1 \tau^i \gamma_2 \tau^j \gamma_3 \tau^k}[A] =
{\textstyle {1\over 4}} W_{\gamma_1 \gamma_3 \bar{\gamma}_2}[A]+
W_{\gamma_2 \gamma_1 \bar{\gamma}_3}[A]+
W_{\gamma_2 \gamma_3 \bar{\gamma}_1}[A].
\end{equation}

It is therefore immediate to find the expression of the determinant of
the metric in the loop representation,
\begin{equation}\label{detg}
\hat{\rm det q} \Psi(\gamma) =
-{\textstyle {1\over 4}}
\epsilon_{abc} \dot{\gamma}_1^a\dot{\gamma}_2^b\dot{\gamma}_3^c \left(
\Psi(\gamma_1 \gamma_3 \bar{\gamma}_2)+
\Psi(\gamma_2 \gamma_1 \bar{\gamma}_3)+
\Psi(\gamma_2 \gamma_3 \bar{\gamma}_1)\right).
\end{equation}

With these elements we are in a position to perform the main calculation
of this paper, to show that the exponential of the Gauss self linking
number of a loop is a solution of the Hamiltonian constraint with a
cosmological constant.

\section{The Gauss (self) linking number as a solution}

In order to be able to apply the expressions we derived in the previous
section for the constraints to the Gauss self linking number we need an
expression for it in terms of which it is possible to compute the loop
derivative. This is furnished by the well known integral expression,
\begin{equation}
{\rm Gauss}(\gamma) = {1\over 4\pi} \oint_\gamma dx^a \oint_\gamma dy^b
\epsilon_{abc} {(x-y)^c \over |x-y|^3}
\end{equation}
where $|x-y|$ is the distance between $x$ and $y$ with a fiducial
metric. This formula is most well known when the two loop integrals are
computed along different loops. In that case the formula gives 1 if the
loops are linked or 0 if the are not. In the present case we are
considering the expression of the linking of a curve with itself. This
is in general not well defined without the introduction of a framing
\cite{GuMaMi}. 

We will rewrite the above expression in a more convenient fashion,
\begin{equation}
{\rm Gauss}(\gamma) = \int d^3x \int d^3y 
 X^a(x,\gamma) X^b(y,\gamma) g_{ab}(x,y)
\end{equation}
where the vector densities $X$ are defined as,
\begin{equation}
X^a(x,\gamma)=\oint_\gamma dz^a \delta(z-x)
\end{equation}
and the quantity $g_{ab}(x,y)$ is the propagator of a Chern-Simons
theory \cite{GuMaMi},
\begin{equation}\label{prop}
g_{ab}(x,y) = \epsilon_{abc} {(x-y)^c \over |x-y|^3}.
\end{equation}

For calculational convenience it is useful to introduce the notation,
\begin{equation}
{\rm Gauss}(\gamma) = X^{ax}(\gamma) X^{by}(\gamma) g_{ax\;by}
\end{equation}
where we have promoted the point dependence in $x$, $y$ to a "continuous
index"
and assumed a "generalized Einstein convention" which means sum over
repeated indices $a$, $b$ and integrate over the three manifold for
repeated continuous indices. This notation is also faithful to the fact
that the index $a$ behaves as a vector density index at the point $x$,
that is, it is natural to pair $a$ and $x$ together.

The only dependence on the loop of the Gauss self linking number is
through the $X's$, so we just need to compute the action of the loop
derivative on one of them to be in a position to perform the
calculation straightforwardly.  In order to do this we apply the
definition of loop derivative, that is, we consider the change in the
$X$ when one appends an infinitesimal loop to the loop $\gamma$ as
illustrated in figure 3.  We partition the integral in a portion going
from the basepoint to the point $z$ where we append the infinitesimal
loop, which we characterize as four segments along the integral curves
of two vector fields $u^a$ and $v^b$ of associated lengths $\epsilon_1$
and $\epsilon_2$ and then we continue from there
back to the basepoint along the loop. Therefore,
\begin{eqnarray}
&&(1+\sigma^{ab} \Delta_{ab}(\gamma_o^z) X^{ax}(\gamma) \equiv
\int_{\gamma_o^z} dy^a \delta(x-y) +
\epsilon_1 u^a \delta(x-z) +\nonumber\\
&&+\epsilon_2 v^b (1+u^c \partial_c)
\delta(x-z) -\epsilon_1 u^a (1+(u^c+v^c) \partial_c) \delta(x-z)-\\
&&-\epsilon_2 v^a (1+v^c \partial_c) \delta(x-z)
+\int_{\gamma_z^o} dy^a \delta(y-z).\nonumber
\end{eqnarray}

\begin{figure}
\epsffile{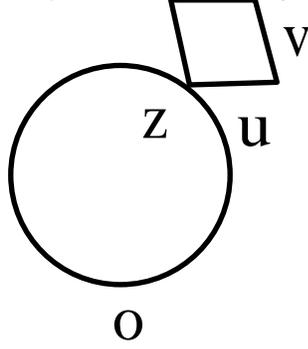}
\caption{The loop derivative that appears in the definition of the
Hamiltonian is evaluated along a path that follows the loop}
\end{figure}
The last and first term combine to give back $X^{ax}(\gamma)$ and
therefore one can read off the action of the loop derivative from the
other terms. Rearranging one geets,
\begin{equation}
\Delta_{ab}(\gamma_o^z) X^{cx}(\gamma) =
\partial_{[a} \delta_{b]}^c(x-z) 
\end{equation}
where the notation $\delta_b^c(x-z)$ stands for $\delta_b^c
\delta(x-z)$, the product of the Kronecker and Dirac deltas.

This is really all we need to compute the action of the Hamiltonian.

We therefore now consider the action of the vacuum ($\Lambda=0$) part of
the Hamiltonian on the exponential of the Gauss Linking number. The
action of the loop derivative is,
\begin{eqnarray}
&&\Delta_{ab}(\gamma_o^x) \exp\left( X^{cy}(\gamma) X^{dz}(\gamma)
g_{cy\;dz} \right)= \\
&&2 \partial_{[a} \delta^c_{b]}(x-y) X^{dz}(\gamma) 
g_{cy\;dz} \exp\left(X^{ew}(\gamma)
X^{fw'}(\gamma) g_{ew\;fw'}\right)\nonumber
\end{eqnarray}
Now we must integrate by parts. Using the fact that $\partial_a
X^{ax}(\gamma) = 0$ and the definition of $g_{cy\;dz}$ (\ref{prop}) we
get
\begin{eqnarray}
&&\Delta_{ab}(\gamma_o^x) \exp\left(X^{cy}(\gamma) X^{dz}(\gamma)
g_{cy\;dz} \right)= \\
&&2 \epsilon_{abc} (X^{cx}(\gamma_1)+
X^{cx}(\gamma_2)+X^{cx}(\gamma_3)
\exp\left(X^{cy}(\gamma) X^{dz}(\gamma)
g_{cy\;dz} \right)\nonumber
\end{eqnarray}

Therefore the action of the Hamiltonian constraint on the Gauss linking
number is,

\begin{equation}
\hat{H} e^{\Lambda {\rm Gauss}} = \epsilon_{abc} \dot{\gamma}_1^a
\dot{\gamma}_2^b \dot{\gamma}_3^c e^{\Lambda {\rm Gauss}}
\end{equation}
where we again have replaced the distributional tangents at the point
of intersection by an expression only involving the tangents. The
expression is only formal since in order to do this a divergent factor
should be kept in front. We assume such factors coming from all terms
to be similar and therefore ignore them.

It is straightforward now to check that applying the determinant of
the metric one the Gauss linking number one gets a contribution
exactly equal and opposite by inspection from expression
(\ref{detg}). This concludes the main proof of this paper.

\section{Discussion}

We showed that the exponential of the Gauss (self) linking number is a
solution of the Hamiltonian constraint of quantum gravity with a
cosmological constant. This naturally can be viewed as the "Abelian"
limit of the solution given by the Kauffman bracket. 

What about the issue of regularization? The proof we presented is only
valid in the limit where $\epsilon \rightarrow 0$, that is, when the
regulator is removed. If one does not take the limit the various terms
do not cancel. However, the expression for the Hamiltonian constraint we
introduced is also only valid when the regulator is removed. A
regularized form of the Hamiltonian constraint in the loop
representation is more complicated than the expression we presented. If
one is to point split, infinitesimal segments of loop should be used to
connect the split points to preserve gauge invariance and a more careful
calculation would be in order. 

What does all this tell us about the
physical relevance of the solutions? The situation is remarkably similar
to the one present in the loop representation of the free Maxwell field
\cite{DiNoGaLeTr}. In that case, as here, there are two terms in the
Hamiltonian that need to be regularized in a different way (in the case
of gravity, the determinant of the metric requires splitting three
points whereas the Hamiltonian only needs two). As a consequence of
this, it is not surprising that the wavefunctions that solve the
constraint have some regularization dependence. In the case of the
Maxwell field the vacuum in the loop representation needs to be
regularized. In fact, its form is exactly the same as that of the
exponential of the Gauss linking number if one replaces the propagator
of the Chern-Simons theory present in the latter by the propagator of
the Maxwell field. This similarity is remarkable. The problematic is
therefore the same, the wavefunctions inherit regularization dependence
since the regulator does not appear as an overall factor of the wave
equation. 

How could these regularization ambiguities be cured? In the Maxwell case
they are solved by considering an "extended" loop representation in
which one allows the quantities $X^{ax}$ to become smooth vector
densities on the manifold without reference to any particular loop
\cite{ArUgGaGrSe}. In the gravitational case such construction is being
actively pursued \cite{DiGaGrPu}, although it is more complicated. It
is in this context that the present solutions really make sense. If one
allows the $X's$ to become smooth functions the framing problem
disappears and one is left with a solution that is a function of vector
fields and only reduces to the Gauss linking number in a very special
(singular) limit. It has been proved that the extension of the Kauffman
Bracket and Jones polynomials to the case of smooth density fields are
solutions of the extended constraints. A similar proof goes through for
the extended Gauss linking number. In the extended representation, there
are additional multivector densities needed in the representation. The
"Abelian" limit of the Kauffman bracket (the Gauss linking number)
appears as the restriction of the "extended" Kauffman bracket to the
case in which higher order multivector densities vanish. It would be
interesting to study if such a limit could be pursued in a systematic
way order by order. It would certainly provide new insights into how to
construct nonperturbative quantum states of the gravitational field.

\section{Acknowledgements}

This paper is based on a talk given by J.P. at the Riverside
conference.  J.P. wishes to thank John Baez for inviting him to
participate in the conference and hospitality in Riverside. This work
was supported by grants NSF-PHY-92-07225, NSF-PHY93-96246, and by
research funds of the University of Utah and The Pennsylvania State
University. Support from PEDECIBA (Uruguay) is also
acknowledged. R.G. wishes to thank Karel Kucha\v{r} and Richard Price
for hospitality at the University of Utah where part of this work was
accomplished.


\begin{thebibliography}{99}
\bibitem{JaSm} T. Jacobson, L. Smolin, Nuc. Phys. {\bf B299}, 295
(1988).
\bibitem{RoSm} C. Rovelli, L. Smolin, Phys. Rev. Lett. {\bf 61}, 1155
(1988); Nuc. Phys. {\bf B331}, 80 (1990).
\bibitem{BrPu} B. Br\"ugmann, J. Pullin, Nuc. Phys. {\bf B363}, 221 (1991).
\bibitem{Ga} R. Gambini, Phys. Lett. {\bf B255}, 180 (1991).
\bibitem{BrGaPu} B. Br\"ugmann, R. Gambini, J. Pullin, Phys. Rev. Lett.
{\bf 68}, 431 (1992).
\bibitem{Wi} E. Witten, Commun. Math. Phys. {\bf 121}, 351 (1989).
\bibitem{GuMaMi} E. Guadagnini, M. Martellini, M. Mintchev, Nuc. Phys.
{\bf B330}, 575 (1990).
\bibitem{BrGaPuessay} B. Br\"ugmann, R. Gambini, J. Pullin, Gen. Rel.
Grav. {\bf 25}, 1 (1993).
\bibitem{Gr} J. Griego, ``The extended loop representation: 
a first approach'', Preprint (1993). 
\bibitem{GaTr} R. Gambini, A. Trias, Phys. Rev. {\bf D27}, 2935
(1983).
\bibitem{DiNoGaLeTr} C. Di Bartolo, F. Nori, R. Gambini, L. Leal, 
A. Trias, Lett. Nuo. Cim. {\bf 38}, 497 (1983).
\bibitem{ArUgGaGrSe} D. Armand-Ugon, R. Gambini, J. Griego, L. Setaro,
``Classical loop actions of gauge theories'', preprint hep-th:9307179 (1993). 
\bibitem{DiGaGrPu} C. Di Bartolo, R. Gambini, J. Griego, J. Pullin, in
preparation. 


\end{thebibliography}
\end{document}